\documentclass[12pt,preprint]{aastex}

\usepackage[colorlinks,
            linkcolor=red,
            anchorcolor=blue,
            citecolor=blue
            ]{hyperref}

\shorttitle{Is flux rope a necessary condition for CMEs?}
\shortauthors{Ouyang, Yang, \& Chen}

\begin{document}

\title{Is flux rope a necessary condition for the progenitor of coronal mass ejections?}

\author{Y. Ouyang\altaffilmark{1,2}, K. Yang\altaffilmark{1,3}, and P. F. Chen\altaffilmark{1,3}}

\affil{$^1$ School of Astronomy \& Space Science, Nanjing University,
        Nanjing 210023, China; \email{chenpf@nju.edu.cn}}
\affil{$^2$ School of Science, Linyi University, Linyi 276000, China}
\affil{$^3$ Key Lab of Modern Astron. \& Astrophys. (Ministry of
        Education), Nanjing University, China}

\begin{abstract}
A magnetic flux rope structure is believed to exist in most coronal mass
ejections (CMEs). However, it has been long debated whether the flux rope exists
before eruption or is formed during eruption via magnetic reconnection. The
controversy has been continuing because of our lack of routine measurements of
the magnetic field in the pre-eruption structure, such as solar filaments.
However, recently an indirect method was proposed to infer the magnetic
field configuration based on the sign of helicity and the bearing direction of
the filament barbs. In this paper, we apply this method to two erupting
filament events, one on 2014 September 2 and the other on 2011 March 7,
and find that the first filament is supported by a magnetic flux rope and the second
filament is supported by a sheared arcade, i.e., the first one is an
inverse-polarity filament and the second one is a normal-polarity filament.
With the identification of the magnetic configurations in these two filaments,
we stress that a flux rope is not a necessary condition for the pre-CME
structure.
\end{abstract}


\section{Introduction}

Coronal mass ejections (CMEs) are the largest-scale and the most intense
eruptions in the solar atmosphere. In each event, a bulk of plasma with mass
up to $10^{14}$--$10^{16}$ g is expelled out from the low corona to the
interplanetary space \citep{webb12}. A CME is manifested in various wavebands
of the electromagnetic spectrum \citep{gopal99}, and the widely used
observations are from white light. After observations for more than 40 years,
the so-called 3-component structure has become the paradigm of CMEs, i.e., a
bright frontal loop, an embedded bright core, and the cavity in between
\citep{ill85}. Calling such a composition the paradigm does not mean that all
CMEs are seen to have all the 3 components, and in fact, many CMEs do not show
all the components, mainly because of the projection effects. The 3-component
structure is often explained in terms of an erupting magnetic flux rope, i.e.,
the piling-up plasma outside the flux rope forms the frontal loop, the
expanding flux rope corresponds to the cavity with an erupting
filament/prominence suspended at the magnetic dips of the flux rope
\citep{forb00}. Alternatively, the frontal loop and the expanding cavity are
explained simultaneously to be due to successive stretching of the magnetic
field lines overlying an erupting flux rope \citep{chen09}. Despite various
possible models for the CME frontal loop and the cavity, it has been well
established that the bright core of the CME corresponds to an erupting
filament/prominence \citep{hous81}.

The bright core is sometimes observed to have a helical structure,
implying the existence of a flux rope \citep{dere99}. After examining the 
Large Angle and Spectrometric Coronagraph
\citep[LASCO,][]{brue95} observations from 1997 to 2010, \citet{vour13} found
that at least 40\% of the CMEs have clear flux rope structures. They even
suggested that flux ropes might be a common structure in CMEs. This is
understandable since no matter a flux rope exists prior to the eruption or not,
magnetic reconnection during the eruption as discussed in the standard CME/flare
model \citep{chen11, shib11} would turn poloidal field into toroidal field,
generating a helical magnetic field, i.e., a flux rope \citep{gosl99}. However,
the ongoing debate is whether the flux rope detected during the eruption phase
exists before eruption or is formed during eruption.

Among the proposed models of CME initiations, some assume that a flux rope
exists before eruption \citep{low95}. In particular, for those models related
to kink or torus instabilities \citep{toro03,kliem06,olme10}, which are found
by \citet{demo10} to be equivalent to the catastrophe model \citep{forb91,
isen93,hu01,lin04,wang15}, the presence of a flux rope is essential. There are
some other models, which assume only a sheared arcade \citep{miki94,anti99} in
their initial conditions. Note that a sheared arcade is not essential in these
models, and can be replaced by a flux rope in principle. However, somehow such
a difference between the indeliberately classified two types of models evolved
into a debate on whether a flux rope is a necessary ingredient of the
progenitor of a CME. As a result, a majority of observational papers tend to
argue that a flux rope exists before eruption, either well before the eruption
\citep{pats13, yan14, fili15} or just prior to eruption \citep{cheng14,kumar14}.
Very few observations supported the scenario of the formation of a flux rope
during deruption \citep[see][for an example]{song14}. However, even based on a
logic test, \citet{chen11,chen12} proposed that flux ropes exist before
eruption in some events and are formed via magnetic reconnection in other
events. The reasoning is very straightforward: It has been well established
that there are both inverse-polarity and normal-polarity filaments, which are
typically described by the Kuperus-Raadu model \citep{kr74} and the
Kippenhahn-Schl\"uter model \citep{ks57}, respectively. In the Kuperus-Raadu
model, a flux rope does exist as also shown by \citet{aula99},
whereas in the Kippenhahn-Schl\"uter model, the filament is supported by a
sheared arcade. Considering that both types of filaments would erupt to form
CMEs, it can be naturally inferred that in the latter case a flux rope, which
might be identified in the extreme ultraviolet (EUV) images or coronagraph
images during eruption, is formed presumably via magnetic reconnection during
eruption.

Systematic measurements of the vector magnetic field in solar prominences were
performed in 1980s \citep{lero84, lero89, bomm98}, which revealed that typically
higher prominences are mainly of the inverse-polarity type, and lower
prominences are mainly of the normal-polarity type. However, despite that
thousands of CMEs have been detected since 1996 thanks to the white-light
observations from the LASCO coronagraph, we have little information about the
magnetic field of the source filaments or prominences of these CMEs because of
the lack of direct measurements of the magnetic field inside these prominences.
Recently, an indirect method to infer the magnetic configuration of solar
filaments was proposed by \citet{chen14}.

Originally, it was believed that there is one-to-one correspondence between the
chirality (hence helicity) of a filament and the bearing of the filament barbs,
i.e., a filament with left-bearing barbs has positive helicity and a filament
with right-bearing barbs has negative helicity \citep[see][for a
review]{mart98}. Hereafter in this paper, this one-to-one correspondence is
called the barb rule. However, \citet{guo10} found that along a filament with
positive helicity, some barbs are left-bearing, and others are right-bearing.
With nonlinear force-free magnetic extrapolations, they demonstrated that the
filament segment following the barb rule is supported by a flux rope, whereas
the filament segment against the barb rule is supported by a sheared arcade.
\citet{chen14} proposed a more general paradigm, claiming that a filament
following the barb rule is hosted by a flux rope, whereas a filament against
the barb rule is hosted by a sheared arcade. Based on this paradigm, once we
know the sign of helicity of a filament, e.g., via vector magnetograms, we can
judge whether the filament is hosted by a flux rope or a sheared arcade by
examining the bearing of the filament barbs in the H$\alpha$ or EUV images.

In this paper, we select two examples of filament eruptions in order to
illustrate that a flux rope is not necessarily the pre-requisite condition for
a CME eruption, i.e., whereas some flux ropes exist prior to eruptions, some
might be formed during CME eruptions via magnetic reconnection. This paper is
organized as follows. The observations are presented in \S\ref{sec2}, the
results are described in \S\ref{sec3}, which are discussed in \S\ref{sec4}.

\section{Observations and Data Analysis}\label{sec2}

\begin{figure}
\epsscale{1.0}
\plotone{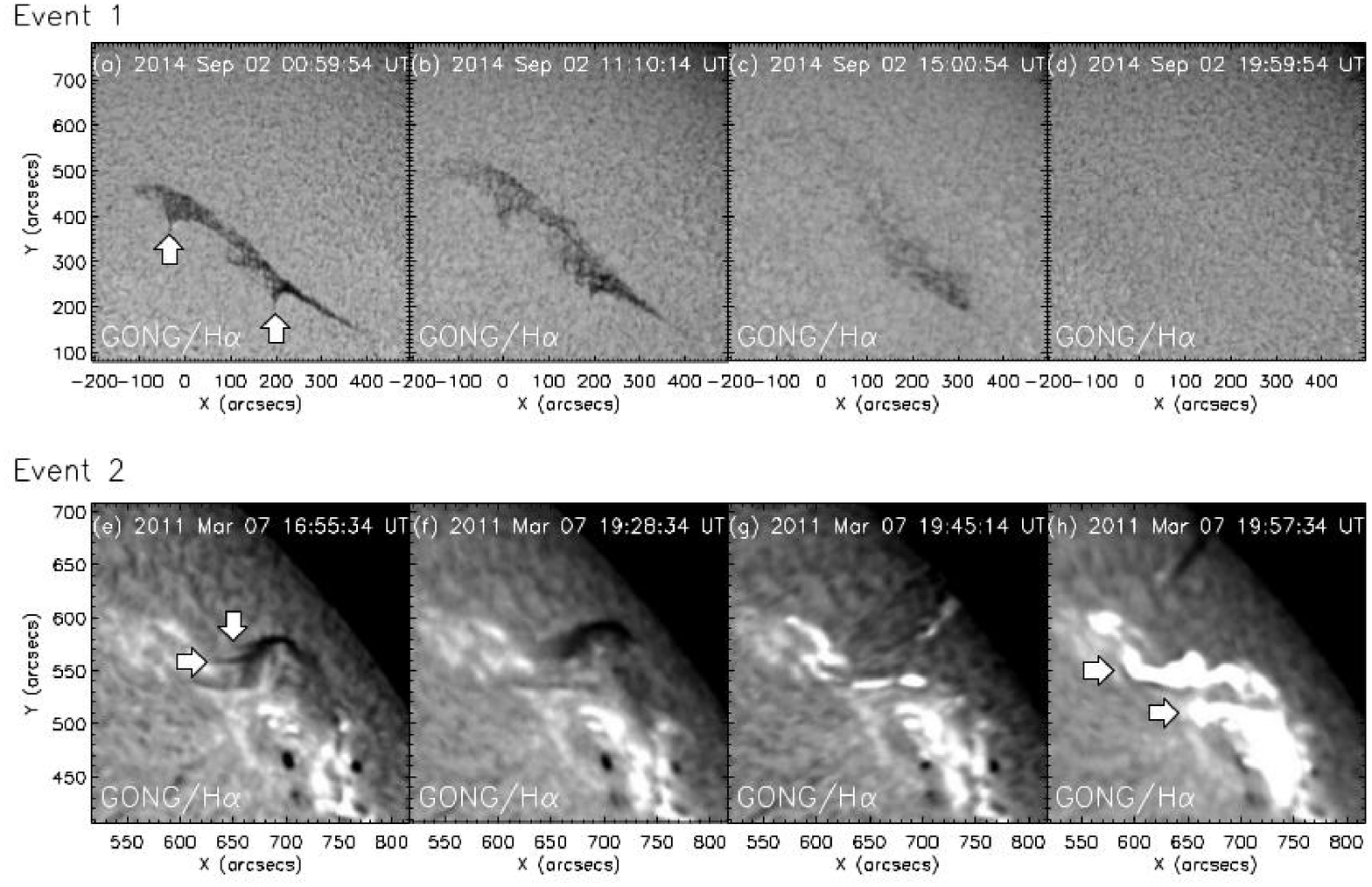}
\caption{Time sequence of the H$\alpha$ images of the two filament eruptions
	 observed by the \emph{GONG} network. {\it Top panels}: the 2014
	September 2 event; {\em Bottom panels}: the 2011 March 7 event.
	\label{fig1}}
\end{figure}

Both filaments under study are located in the northern hemisphere, and both are
associated with a solar flare and a CME. The first event, event 1 hereafter, is
a candidate filament with the inverse-polarity magnetic configuration, which
erupts on 2014 September 2. As shown by the top panels of Figure \ref{fig1},
the filament is rather elongated, extending more than 600\arcsec\ in the quiet
region. Its eruption leads to a partial halo CME and a {\it GOES} C2.5-class
flare. The corresponding Solar Object Locater is
SOL2014-09-02T11:10:03L205C040. The second event, event 2 hereafter, is a
candidate filament with the normal-polarity magnetic configuration, which erupts
on 2011 March 7. As displayed by the bottom panels of Figure \ref{fig1}, this
filament is located in the active region NOAA 11164, curving like an arch bridge
along the magnetic polarity inversion line (PIL). It belongs to the active
region type \citep{mack10}. The eruption is accompanied by a {\it GOES}
M3.7-class flare and a halo CME as well. The corresponding Solar Object Locater
is SOL2011-03-07T19:00:01L193C050.

Both eruptions are well observed by the ground-based and space-borne
telescopes. The evolution of the filaments is clearly monitored by the {\it
Global Oscillation Network Group} ({\it GONG}) in H$\alpha$ \citep{harv11},
and is also observed in EUV images obtained by the Atmospheric Imaging Assembly
\citep[AIA,][]{leme12} aboard the {\em Solar Dynamics Observatory} ({\it SDO}).
The Helioseismic and Magnetic Imager \citep[HMI,][]{sche12} aboard {\it SDO}
provides the photospheric vector magnetograms below the erupting filament. The
corresponding CMEs are registered by LASCO coronagraph on board the {\it Solar
and Heliospheric Observatory} ({\it SOHO}) spacecraft. The {\it SDO}/AIA
observes the full solar disk with a pixel size of $0\farcs 6$ and a time
cadence of $\sim$12 s in three UV and seven EUV passbands. In this paper, we
use the 193 \AA\ and 304 \AA\ passbands, in addition to the
{\it GONG}/H$\alpha$ images, to trace the evolution of the erupting filament
events. For event 1, all the  disk images are derotated to the universal time
00:00:00 UT on the day of eruption; for event 2, all the disk images are
derotated to 21:30:00 UT on the day of eruption.

\section{Results}\label{sec3}

\subsection{Event 1} \label{evt1}

On 2014 September 2, a quiescent filament is situated
 across the central meridian in the northern hemisphere, appearing with an inverse-S shape. As
shown in the top panels of Figure \ref{fig1}, two right-bearing barbs can be
cleared identified at $\sim$00:59 UT well before the slowly ascending phase of
the filament, as indicated by the white arrows. At $\sim$08:00 UT, the northern
part of the elongated filament starts to rise slowly, and the southern part
follows. The eruption is so slow that the rising filament is discernable in
H$\alpha$ for more than 7 hr. At $\sim$15:00 UT, a major part of the filament 
vanishes from the H$\alpha$ image. As time goes on, the whole filament 
disappears. However, no clear flare ribbons are discernable in H$\alpha$.

\begin{figure}
\epsscale{1.0}
\plotone{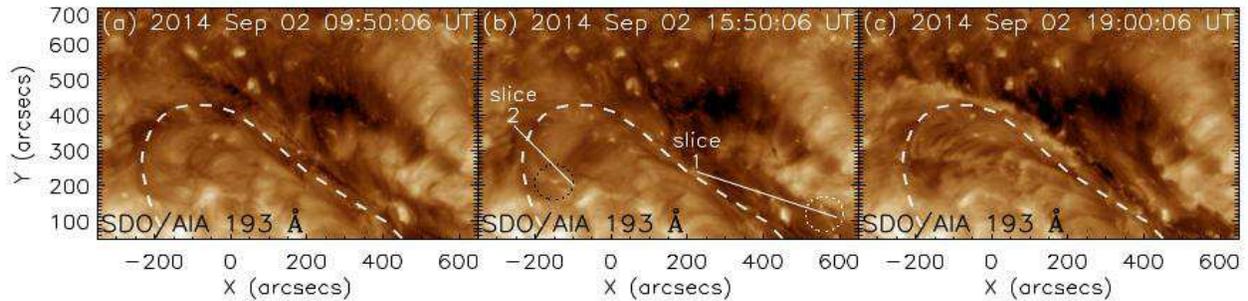}
\caption{Time evolution of the filament activation on 2014 September 2
	observed by \emph{SDO}/AIA 193 \AA\ channel. The black and white
        circles in panel (b) mark the brightenings where the draining filament
	plasmas impact the solar surface. The white solid lines 1 and 2 in
	panel (b) mark the slices for plotting the time-slice diagrams in Fig.
	\ref{fig3}. The two conjugate drainage sites are left-skewed compared
	to the underlying magnetic polarity inversion line, which is marked by
	the thick dashed lines.}
  \label{fig2}
\end{figure}

\begin{figure}
\epsscale{1.0}
\plotone{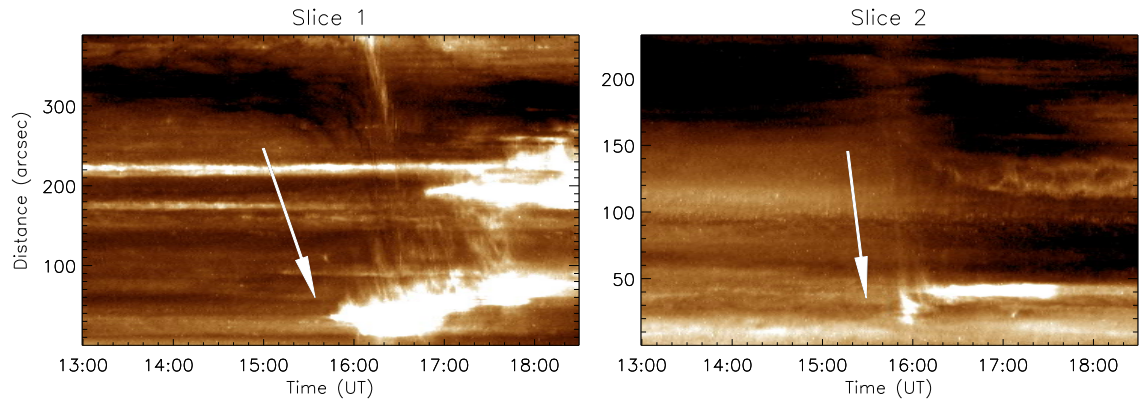}
\caption{Time-distance diagrams of the AIA 193 \AA\ intensity along the slices 1
	({\it left}) and 2 ({\it right}), showing the plasma drainage along the
	two legs of the filament on 2014 September 2. The locations of the two
	slices are marked in Fig. \ref{fig2}(b). The start points of the two
	slices are both at the western ends.}
  \label{fig3}
\end{figure}

The activation of the filament is better seen from the high-resolution images
of the {\it SDO}/AIA. The time sequence of the 193 \AA\ images is displayed in
Figure \ref{fig2} and the associated animation. It is seen that as the filament
starts to rise, its threads become more and more dynamic.  As seen in many
other erupting filaments, the ascending of the main body of the filament is
accompanied by the drainage of the cool material near the two ends, presumably
sliding along the embedded magnetic field lines. At the eastern end, the
drainage plasma changes from absorption to emission. After moving along a
curved trajectory in an anti-clockwise direction, the draining plasma impacts
the base of the corona, forming a bright spot as indicated by the black circle
in Figure \ref{fig2}(b). Almost simultaneously, a big bundle of plasma is seen
to fall down near the western end of the filament and hits the coronal base,
forming a narrow band of brightening as indicated by the white circle in
Figure \ref{fig2}(b). In order to see the draining motions clearly, we select
two slices near the two ends of the filament respectively, which are marked by
the white solid lines in Figure \ref{fig2}(b). The time-slice diagrams of the
AIA 193 \AA\ intensity are depicted in Figure \ref{fig3}, where the start points
of the two slices are at the western end in both cases. The draining motion of
the filament is indicated by the white arrows. It is seen that near the two
ends of the filament, plasma drains down and then impacts the solar surface,
forming brightenings. It is also noticed that the draining plasma is already
heated before the impact, as revealed by the bright threads in Figure
\ref{fig3}. By comparing the line connecting the two conjugate drainage sites
with the magnetic PIL ({\it dotted line}) in Figure \ref{fig2}, it is found
that the drainage sites are left-skewed according to the definition in
\citet{chen14}.

\begin{figure}
\epsscale{1.0}
\plotone{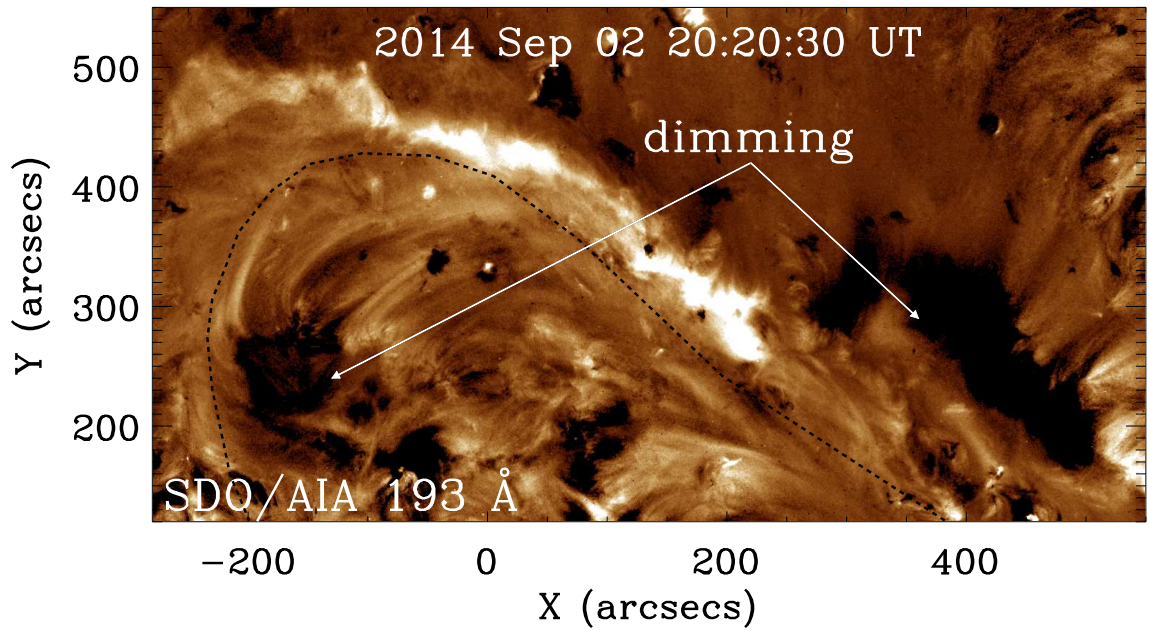}
\caption{EUV difference image at 20:20:30 UT in the aftermath of the filament
	eruption on 2014 September 2 observed by \emph{SDO}/AIA in 193 \AA\
	showing twin dimmings and flare loops. The base image at 11:40:06 UT
	is subtracted. It is seen that the twin dimmings are left-skewed
	compared to the magnetic PIL which is marked by the dotted line.}
  \label{fig4}
\end{figure}

As the filament erupts, both the embedded and the overlying magnetic fields are
expected to expand as well, forming twin dimmings near the source region
\citep{ster97, harr03}. Figure \ref{fig4} displays the {\it SDO}/AIA 193 \AA\
base-difference image at 20:20:30 UT,  which is subtracted with the
pre-eruption image at 11:40:06 UT. A pair of dimmings is seen to be located on
the two sides of the magnetic PIL. It is obvious that the twin dimmings are
left-skewed compared to the magnetic PIL. It is noted that flaring loops are
visible in the 193 \AA\ image in Figure \ref{fig4}, though the northern flare
ribbon is much brighter than the southern one. Besides, Figure \ref{fig4}
reveals that the flaring loops are also left-skewed compared to the magnetic
PIL.

\begin{figure}
\epsscale{1.0}
\plotone{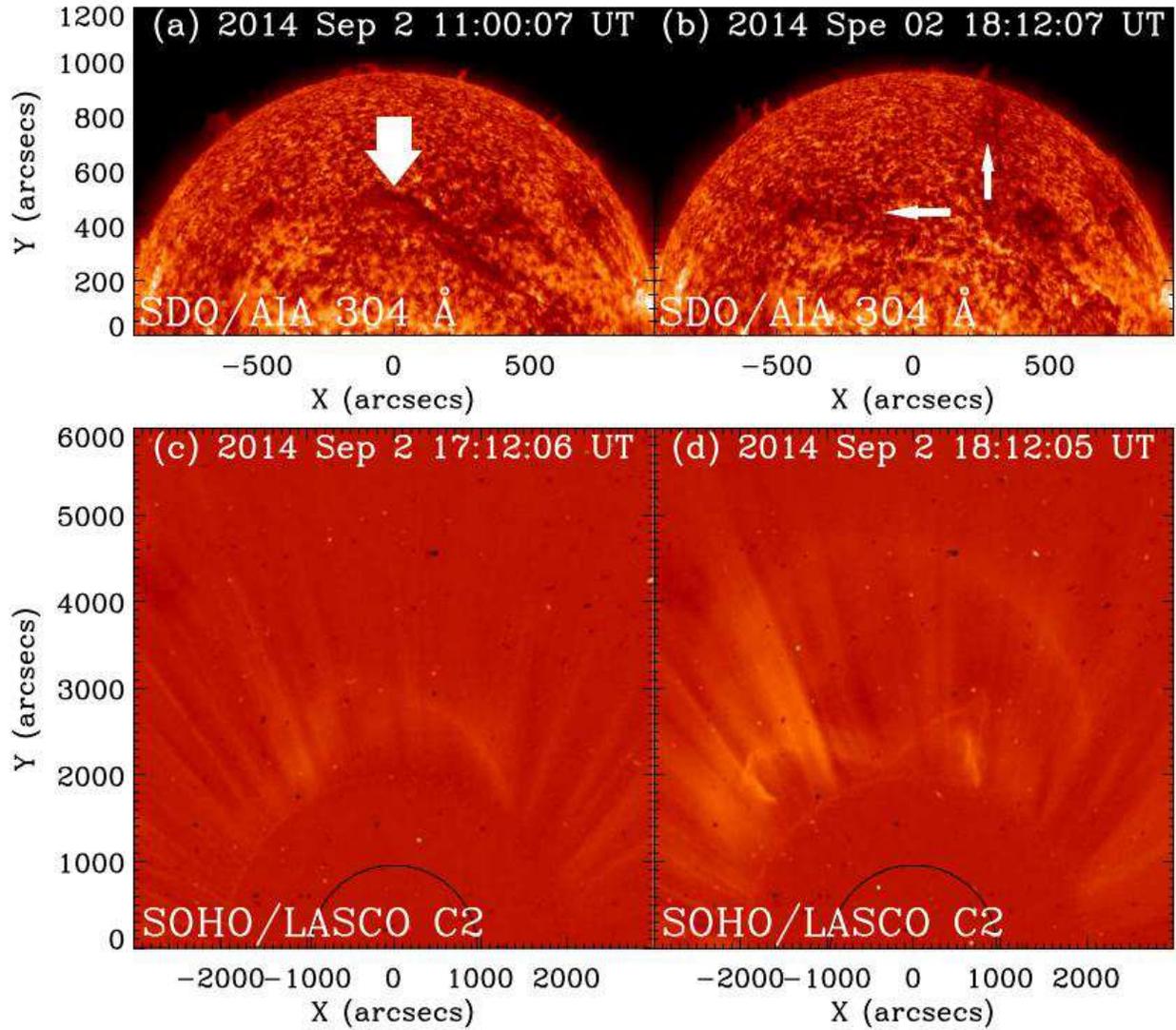}
\caption{Evolution of the erupting filament observed by \emph{SDO}/AIA in the
	304 \AA\ channel ({\it top}) and the associated CME observed by the
	\emph{SOHO}/LASCO coronagragh ({\it bottom}) on 2014 September 2. In
	the top panels, the filament is indicated by the arrows. In the bottom
	panels, the black circle marks the solar limb.}
  \label{fig5}
\end{figure}

Figure \ref{fig5} shows the filament eruption observed by \emph{SDO}/AIA in 304
\AA\ and the accompanied CME observed by the LASCO white-light coronagraph. It
is seen that the filament initially resides in the source region (panel a), and
rises to a large distance from the source region at 18:12:07 UT as indicated by
the white arrow in panel (b). At 17:12:06 UT, the corresponding CME frontal loop
first appears in the LASCO C2 field of view, whereas at 18:12:05 UT all the 3
components of the CME, i.e., a frontal loop, a bright core, and a cavity in
between, are clearly visible. We compare the position angles of the eastern and
western legs of the erupting filament in panel (b) and of the two legs of the
CME core in panel (d), and find that the position angles of the filament legs
are only $7\pm2^\circ$ larger than that of the CME core. Considering the
curved structure of the erupting filament, it would be fair to claim that the
bright CME core indeed corresponds to the erupting filament.

\subsection{Event 2}

On 2011 March 7, a thick filament is located inside the active region NOAA
11164 in the northern hemisphere, hence it is an active region filament. As
shown in panel (e) of Figure \ref{fig1}, the filament looks like an arch
bridge, curving with a forward S-shape. Its activation and the ensuing solar
flare are observed by {\it SDO}/AIA with high spatiotemporal resolution. A time
sequence of the 304 \AA\ images is displayed in Figure \ref{fig6} and the
associated animation. Before eruption, the filament is seen as a cluster of
threads above the magnetic PIL as shown by Figure \ref{fig6}(a) and its inset.
Compared to the the filament spine or the underlying magnetic PIL as marked by
the dotted line in Figure \ref{fig6}, the filament threads are right-skewed.
The right-skewed barbs are also clearly visible in the H$\alpha$ images as
shown in Figure \ref{fig1}(e). It should be noted that barbs are 
a dynamic structure, forming and disappearing from time to time \citep{li13},
and the H$\alpha$ barbs are not always visible. In this case, we can judge
the bearing sense of a filament based on the threads. Actually, threads are
even more reliable than H$\alpha$ barbs, as demonstrated by \citet{mart08}.

At $\sim$19:20 UT, a brightening starts to appear around the western end of the
filament. With the brightening increasing in strength and extending eastward.
The central part of the filament spine begins to brighten. With that, the
filament is broken into two parts. As the western part falls down, the eastern
part rises up rapidly.  At $\sim$19:45 UT, flare ribbons and flaring loops
become visible. As the eruption continues, erupts, a small bundle of plasma in
absorption is seen to fall down near the western end of the filament at
$\sim$19:36:08 UT, forming a bright narrow band at the edge of the draining
plasma as enclosed by the white circle in Figure \ref{fig6}(b). Soon after, a
large amount of filament material falls down to the solar surface along the
eastern leg of the filament, forming another drainage site near the eastern
end, which is indicated by the black circle in Figure \ref{fig6}(b). 
Similarly,
we select two slices near the two ends of the filament to show the the draining
motions, respectively. The two slices are marked by the white solid lines 1 and
2 in Figure \ref{fig6}(b). The time-slice diagrams of the AIA 304 \AA\ intensity
are depicted in Figure \ref{fig7}, where the start point of line 1 is the
southern end and that of line 2 is the eastern end. The draining motions and the
brightening are clearly revealed in the figure, as indicated by Figure
\ref{fig7}. By comparing the line connecting the two conjugate drainage sites
with the magnetic PIL, it is found that the drainage sites are right-skewed
according to the definition in \citet{chen14}. The flaring loops are also seen
to be right-skewed, as revealed by Figure \ref{fig6}(c).

Considering that the bearing sense of the filament barbs and the skew of the
flaring loops may suffer from the projection effects since the filament is
close to the solar limb, we examine the H$\alpha$ and EUV images at $\sim$06:12
UT on 2011 March 4 when the filament was passing through the central meridian.
As shown by both the  H$\alpha$ and AIA 193 \AA\ images in Figure \ref{fig8},
the filament appears with an `S' shape with right-bearing barbs as indicated by
the thick arrows, and the overlying coronal arcade is right-skewed in
comparison with the filament spine, as indicated by the thin arrows. These
results confirm that the filament indeed has right-bearing barbs, and the
overlying coronal magnetic loops are right-skewed.

In order to check the skew of the EUV twin dimmings, we display the AIA 193
\AA\ difference image at 21:29:08 UT in Figure \ref{fig9}, which is subtracted
with the pre-eruption image at 17:39:08 UT. The twin dimmings are clearly seen
to be right-skewed, as indicated by the two arrows. In this figure, the bright
flaring loops are also revealed to be right-skewed when compared with the
underlying magnetic PIL, consistent with Figures \ref{fig6} and \ref{fig8}.

Figure \ref{fig10} depicts the filament eruption observed by {\em SDO}/AIA in
304 \AA\ and the accompanied CME by the LASCO white-light coronagraph. It is
seen through the movie that the filament starts to accelerate significantly at
$\sim$19:44 UT. At 20:12:06 UT when it is close to the flare peak in {\em GOES}
1--8 \AA, both the CME frontal loop and the bright core are discernable, but
the CME cavity is not clear. At 20:24:06 UT, all the 3 components of the CME
are visible and the CME evolves into a halo event. We compare the position
angle of the northern leg of the erupting filament in panel (b) and that of the
northern leg of the CME core in panel (d), and find that they are nearly the
same, implying that the bright CME core indeed corresponds to the erupting
filament.

\section{Discussions}\label{sec4}

\subsection{Helicity of the two filaments}

Magnetic field in the solar corona is often sheared or twisted, which offers
free energy for solar flares and CMEs \citep{chen11,low15}. The magnetic twist
or shear can be quantitatively measured by helicity, either magnetic helicity
or current helicity \citep{demo97}. It has been found that there is a
preferential sign of current helicity in each hemisphere, i.e., negative in the
northern hemisphere and positive in the southern hemisphere. Such a preference
is valid not only for active regions \citep{seeh90, pevt94, rust94,bao98}, but
also for solar filaments \citep{pevt03}. Generally it is assumed that current
helicity and magnetic helicity have the same sign, especially for the direct
current system \citep{demo97}. Since it is much easier to calculate the current
helicity density, it is convenient to use the twist parameter $\alpha$ to
represent the helicity of a magnetic system \citep[e.g.,][]{seeh90}, where
$\alpha$ is related to the magnetic field by the formula $\nabla \times
B=\alpha B$ under the assumption  of the force-free condition in the
photosphere. With photospheric vector magnetograms, we can get the vertical
components of both the magnetic field ($B_z$) and the current density ($J_z$)
in the local Cartesian coordinates. It is noted in passing that even the full
vector of the current density can also be derived under the force-free
assumption \citep{yang14}. Thus, we can calculate the helicity parameter
$\alpha$ via $\alpha=(\nabla \times B)_z/B_z$, i.e.,

\begin{equation}\label{eq1}
\alpha= ( \bigtriangledown \times \textbf{B})_{z} / B_{z} = (\partial B_{x} / \partial y - \partial B_{y} / \partial x) / B_{z},
\end{equation}

\noindent
so as to determine the sign of helicity of the filaments studied in this paper.
A positive/negative $\alpha$ corresponds to a sinistral/dextral filament.

In the case without vector magnetograms, we may have other indirect methods to
infer the sign of helicity of a magnetic system, e.g., the sunspot whorls
\citep{zirk97}. In terms of solar filaments, several indirect methods have been
proposed to infer the sign of helicity, e.g., the bearing of the filament barbs
\citep{mart94,mart98}, overlying coronal arcades \citep{mart95}, EUV twin
dimmings \citep{jiang11}, and the skew of the conjugate drainage sites upon
filament eruptions \citep{chen14}. According to these methods, a dextral
filament would have right-bearing barbs, left-skewed coronal arcades, and
left-skewed twin dimmings and drainage sites upon eruption, whereas a sinistral
filament would have left-bearing barbs, right-skewed coronal arcades, and
right-skewed twin dimmings and drainage sites upon eruption. If confirmed,
these methods may provide an efficient way to determine the sign of helicity of
a filament and the corresponding CME after the filament erupts. Since the sign
of helicity of CMEs is one important parameter to determine the
geo-effectiveness of the CME, and CMEs often have the same sign of helicity as
the source filaments \citep{jing04}, the quick determination of the sign of
helicity for solar filaments will be beneficial for space weather forecasting.

According to our data analysis in \S\ref{sec3}, it is found that for the 2014
September 2 event, the filament barbs are right-bearing, the twin dimmings are
left-skewed, the flaring loops are left-skewed, and the conjugate sites of the
filament drainage are left-skewed as well. Note that the skew of all these
patterns is judged when compared with the photospheric magnetic PIL or the
filament spine. Based on the rules mentioned above, this filament has negative
helicity, which is the preferential helicity in the northern hemisphere where
the filament is located. In order to check the validity of these indirect
methods, we calculate the twist parameter $\alpha$ averaged around the filament 
channel using the vector magnetograms measured by {\it SDO}/HMI. The 
area of calculation is marked as the white box in the left panel of Figure 
\ref{fig11}, which shows the longitudinal magnetogram in gray-scale. After 
removing the $180^\circ$ ambiguity and the projection effects, we use Eq. 
(\ref{eq1}) to calculate $\alpha$ at each pixel inside the box. Considering the 
singularity of Eq. (\ref{eq1}), we ignore those pixels with $|B_z|\leq 10$ G in 
order to reduce the error. The averaged $\alpha$ in the selected box is -0.078 
Mm$^{-1}$. As the box becomes wider, the averaged $\alpha$ changes in the range 
of -0.01 and -0.08 Mm$^{-1}$, meaning that the filament channel has a negative 
helicity. Such a result confirms that all the indirect methods mentioned above 
succeed in inferring the sign of helicity of the erupting filament event on 
2014 September 2.

For the 2011 March 7 event, it is found according to our data analysis in
\S\ref{sec3} that the twin dimmings are right-skewed, the flaring loops are
right-skewed, and the conjugate sites of filament drainage are right-skewed as
well. These features all imply that the helicity of this filament has a
positive sign. However, the filament barbs are revealed to be right-bearing.
If we apply the barb rule mentioned in \citet{mart98}, we would claim that the
helicity of this filament is negative, which is opposite to the results
inferred from all other indirect methods. In order to check its real sign of
helicity, we use the {\it SDO}/HMI vector magnetograms to calculate the twist
parameter $\alpha$. With the same method as above, we calculate the 
averaged $\alpha$ for the 2011 March 7 event. In order to avoid the projection 
effects, the magnetogram of the corresponding active region on March 4 is used, 
as shown in the right panel of Figure \ref{fig11}. The averaged $\alpha$ inside 
the white box is $\sim$0.39 Mm$^{-1}$, which is nearly independent of the size 
of the selected area. We also exclude the sunspot patch and consider those 
pixels with $10\leq |B_z|\leq 100$ G only, the resulting $\alpha$ is 
$\sim$0.40 Mm$^{-1}$. This means  that the erupting filament on 2011 March 7 
has a positive helicity, which is the same as the active region.
Note that the result is opposite to the preferential
sign of helicity in the northern hemisphere where the filament is located. Such
a result validates again the indirect methods based on the skew of the flaring
loops \citep{mart95}, the skew of the twin dimmings \citep{jiang11}, and the
skew of the conjugate sites of the filament drainage \citep{chen14}. However,
it contradicts the result derived from the filament barbs under the assumption
that there is one-to-one correspondence between the filament chirality and the
barb bearing.

The reason of this contradiction is that, as suggested by \citet{chen14}, the
one-to-one correspondence between the filament chirality and the barb bearing
reviewed by \citet{mart98}, i.e., a sinistral/dextral filament has
left/right-bearing barbs, is valid only for the inverse-polarity filaments,
i.e., those
with a flux rope magnetic configuration. For normal-polarity filaments, i.e.,
those with a sheared arcade magnetic configuration, the correspondence is
opposite, i.e., dextral filaments have left-bearing barbs, and sinistral
filaments have right-bearing barbs. According to the paradigm proposed by
\citet[][see their Fig. 7]{chen14}, the sinistral chirality of the 2011 March
7 filament, with a positive helicity, combined with its right-bearing barbs,
implies that the corresponding magnetic structure of the 2011 March 7 filament
is a sheared arcade, rather than a flux rope. The hot channel
structure observed during the eruption of this filament, which was proposed to
be the evidence of a flux rope by \citet{cheng13}, should be formed during the
initiation phase of the eruption, e.g., via magnetic reconnection.

\subsection{Is flux rope a necessary condition for the CME progenitor?}

It is well known that most CMEs derive their energy from the hosting magnetic
field \citep{forb00, chen11}. Therefore, the pre-eruption structure, sometimes
called CME progenitor \citep{chen11}, becomes a key issue to explore in the CME
research. Since flux rope structures were widely identified to exist in CMEs
or interplanetary CMEs \citep[ICMEs,][]{burl81, cheng13, vou14}, a natural
question is
whether the magnetic flux rope exists prior to eruption, i.e., as a CME
progenitor, or it is formed during eruption. A flux rope is an ideal candidate
for the CME progenitor since it represents twisted magnetic fields and hence
contains free magnetic energy in order to power CMEs and solar flares. Besides,
it is easy for a magnetic structure with a flux rope to lose its equilibrium
\citep{forb93} and be triggered to erupt through various types of instabilities
\citep{toro03,kliem06,olme10}. From the observational point of view, many case
studies have already shown that a magnetic flux rope exists before eruption.
Whereas these authors presented convincing evidence of a pre-existing flux rope
in their observations, we still cannot exclude the possibility of a CME
progenitor without a flux rope.

\begin{figure}
\epsscale{1.0}
\plotone{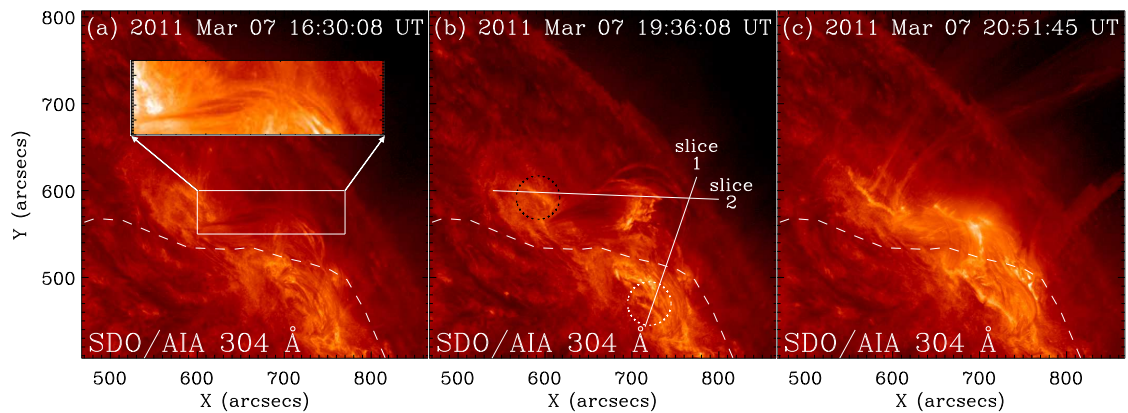}
\caption{Time evolution of the filament activation on 2011 March 7 observed
	by the \emph{SDO}/AIA 304 \AA\ channel, where the dotted lines
	indicate the magnetic PIL. The inset in panel (a) highlights the
	threads in a segment of the filament, where right-bearing barbs can
	be seen. The white and black circles in panel (b) mark the drainage
        sites where the filament plasmas impact the solar surface. The two
        conjugate drainage sites are right-skewed compared to the magnetic PIL.
        The white solid lines 1 and 2 in panel (b) mark the slices for plotting
	the time-slice diagrams in Fig. \ref{fig7}.}
  \label{fig6}
\end{figure}

\begin{figure}
\epsscale{1.0}
\plotone{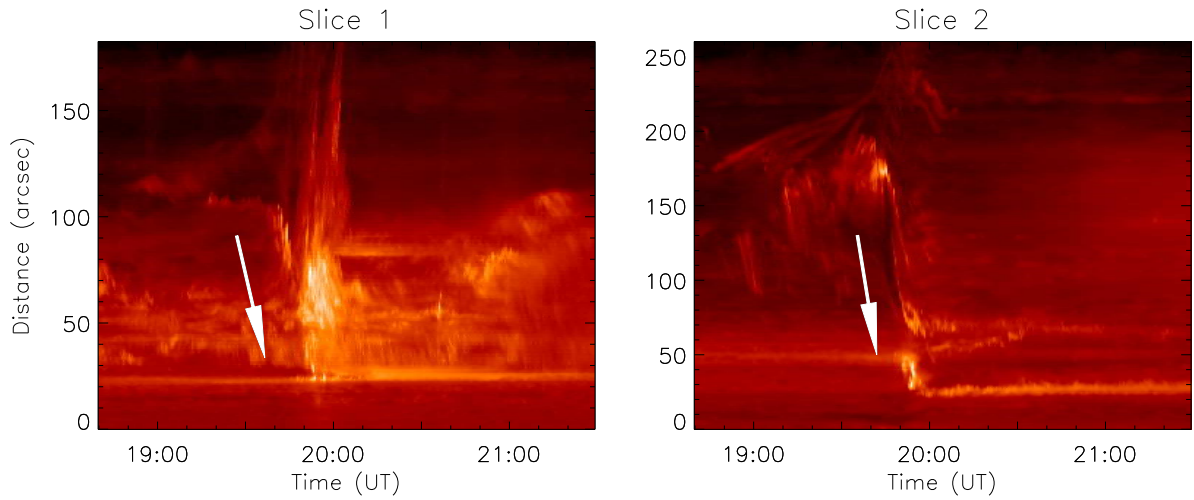}
\caption{Time-distance diagrams of the AIA 304 \AA\ intensity along the slices 1
        ({\it left}) and 2 ({\it right}), showing the plasma drainage along the
	two legs of the filament on 2011 March 7. The locations of the two
        slices are marked in Fig. \ref{fig6}(b). The start point is 
	at the southern end for slice 1 and the eastern end for slice 2. }
  \label{fig7}
\end{figure}

\begin{figure}
\epsscale{1.0}
\plotone{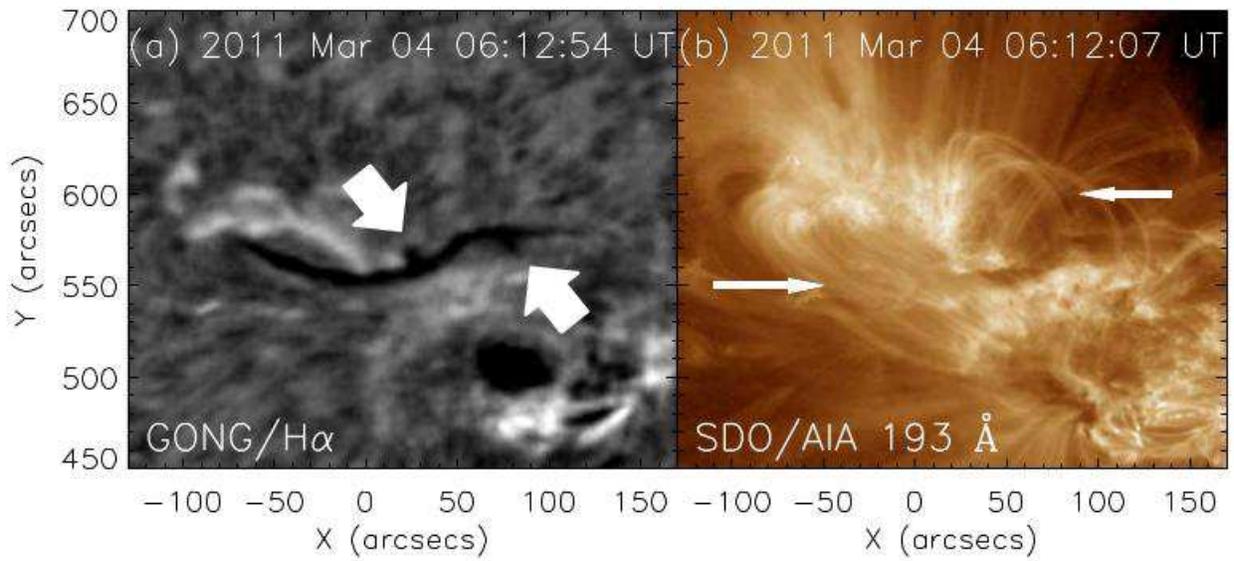}
\caption{The \emph{GONG} H$\alpha$ image ({\it left}) and the \emph{SDO}/AIA
	193 \AA\ image ({\it right}) of the filament in event 2 observed 3
	days before eruption when the filament passes the central meridian.
	Right-bearing barbs are seen as indicated by the thick arrows, and
	the right-skewed coronal arcade is discernable as indicated by the thin
	arrows.}
  \label{fig8}
\end{figure}

\begin{figure}
\epsscale{1.0}
\plotone{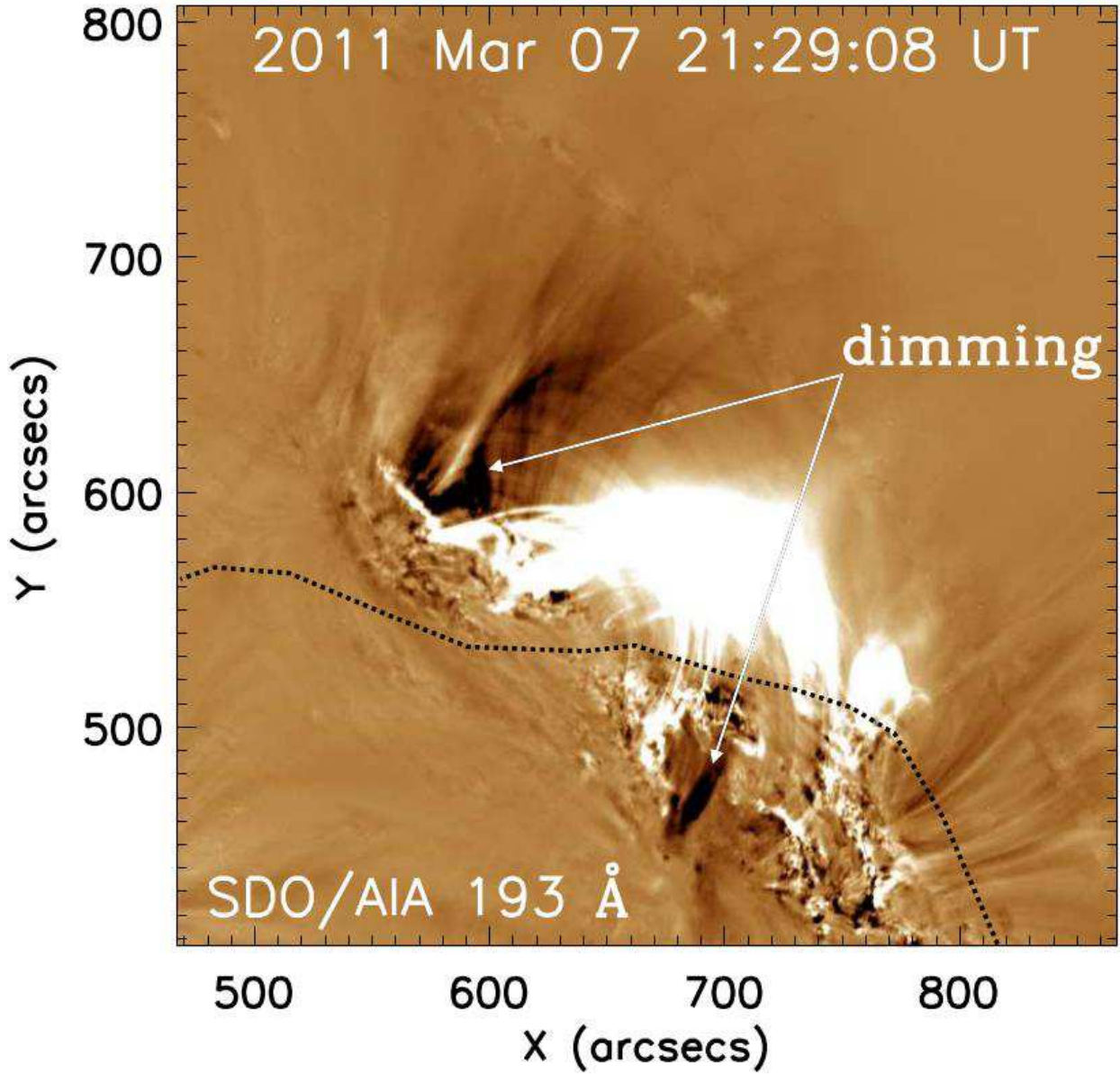}
\caption{EUV difference image of the filament eruption event at 21:29:08 UT
	on 2011 March 7 observed by \emph{SDO}/AIA in 193 \AA\ showing twin
	dimmings and flare loops in the aftermath of the eruption. The base
	image at 17:39:08 UT is subtracted. It is seen that the twin dimmings
	are right-skewed compared to the magnetic PIL which is marked by the
	dotted line.}
  \label{fig9}
\end{figure}

\begin{figure}
\epsscale{1.0}
\plotone{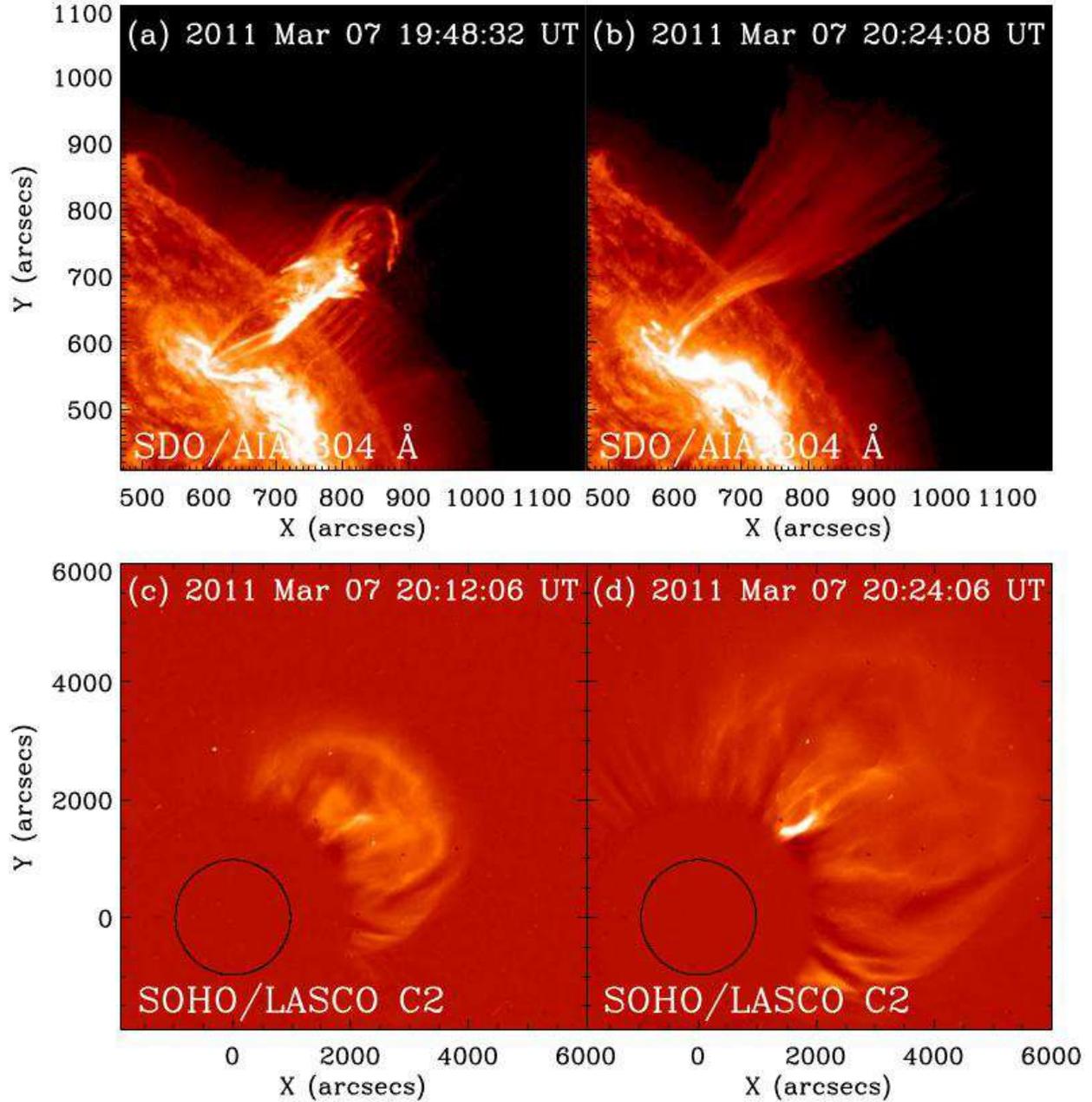}
\caption{Evolution of the erupting filament observed by \emph{SDO}/AIA in
	304 \AA\ ({\it top}) and the associated CME observed by the
	\emph{SOHO}/LASCO coronagraph ({\it bottom}) on 2011 March 7. In the
	bottom panels, both the CME frontal loop and the bright core are
	visible, where the black circle is the solar limb.}
  \label{fig10}
\end{figure}

\begin{figure}
\epsscale{1.0}
\plotone{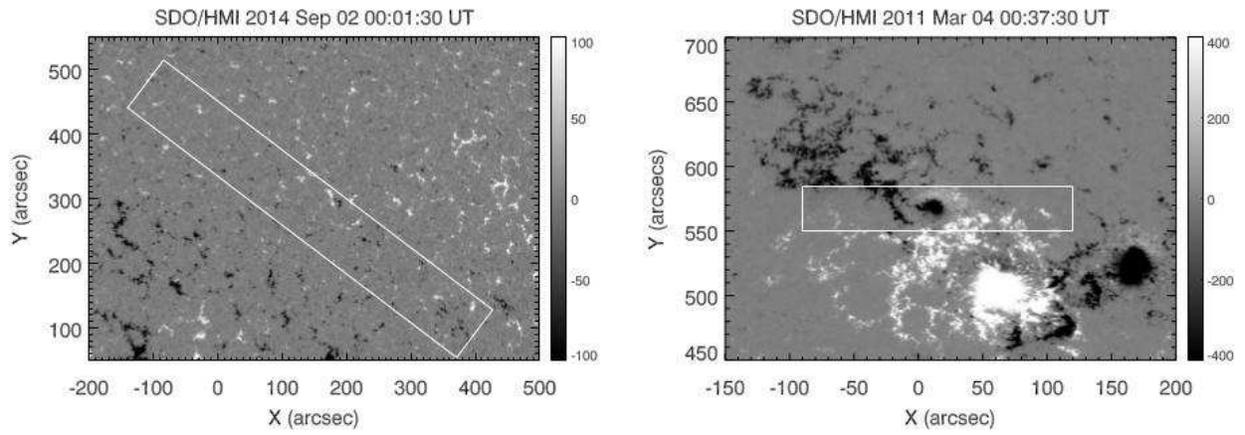}
\caption{Longitudinal magnetograms of the two filament eruption events
	observed by \emph{SDO}/HMI, where the white boxes are selected areas
	for the calculation of the average twist parameter $\alpha$. The 
	magnetic field in the color bars is in units of G.}
  \label{fig11}
\end{figure}

From another line of thought, it is well known that many filaments end up with
eruptions, forming the core of CMEs \citep{hous83, schm13}. Polarization 
measurements have revealed that there are both inverse-polarity type and 
normal-polarity type of filaments \citep{lero84, lero89, bomm98}. 
Correspondingly, they are described by the flux rope model \citep{kr74} and the
sheared arcade model \citep{ks57, aula00}, respectively. Based on a logic test,
\citet{chen11, chen12} proposed that a flux rope is not a necessary condition 
for a CME progenitor, and a sheared arcade can also erupt to form a CME. In 
this case, the strongly sheared core field along the filament channel may not 
possess those ideal MHD instabilities found in the configuration with a flux 
rope, it may still erupt when it is triggered by magnetic breakout above the 
core field \citep{anti99} or emerging flux inside or outside the filament 
channel \citep{feyn95, chen00}. Besides, it would be expected to see a 
quasi-separatrix layer along the strongly sheared core field, where magnetic 
reconnection would play the same role as tether-cutting \citep{moor01}, 
triggering the filament to erupt. The last mechanism probably works for
the 2011 March 7 filament eruption event since brightenings are seen in the 
central part of the
filament at $\sim$19:20 UT, as evidenced in the attached movie of Figure
\ref{fig6}. Such brightenings are indicative of magnetic reconnection in the
strongly sheared arcade field. Interestingly, \citet{cheng13} analyzed the
EUV observations of this event carefully and found that a hot channel, which 
was interpreted to be a flux rope, starts to be visible also at $\sim$19:20 
UT. Our results indicate that it is magnetic reconnection that changes a 
sheared arcade into a flux rope during the CME initiation process.

Unfortunately, there are no routine magnetic field measurements for solar
filaments, except sparse individual ones \citep[e.g.,][]{kuck09, xu12, oroz14,
sass14}. However, the paradigm presented in Figure 7 of \citet{chen14} offers
an indirect method to diagnose the magnetic configuration of a filament by
combining the information of the sign of helicity and the barb bearing of
the filament, i.e., a sinistral filament with left-bearing barbs should have
a flux rope magnetic configuration, and a sinistral filament with right-bearing
barbs should have a sheared arcade magnetic configuration. For dextral
filaments, the correspondence is opposite. Note that a filament may have barbs
with mixed bearings. In this case the dominating bearing is considered
\citep{pevt03, hao15}. Applying this model to the erupting filaments on 2014
September 2 and 2011 March 7, we conclude that the former filament has a
magnetic flux rope before eruption, and the latter filament has a sheared
arcade magnetic configuration before eruption. With these two cases
representing two types of magnetic configurations of solar filaments, i.e.,
inverse-polarity and normal-polarity filaments, we stress that a flux rope is
not a necessary condition as a pre-eruption structure leading to a CME. Of
course, as a filament rises, no matter with a flux rope or a sheared arcade,
the ensuing magnetic reconnection below the core field would turn poloidal
magnetic field to toroidal field \citep{qiu07, hu14}, forming a new flux rope
or growing the pre-existing flux rope. This is why the 3-component structure,
which implies the existence of a flux rope, was observed in both CME eruptions.

\acknowledgments
The authors thank the {\em GONG}, {\em SDO}, and {\em SOHO} teams for making
the data publicly available. This research was supported by the Chinese
foundations 2011CB811402, NSFC (11025314 and 11533005), and Jiangsu 333 Project.

\clearpage

\clearpage

\clearpage

\clearpage

\clearpage

\clearpage

\clearpage

\clearpage

\clearpage

\clearpage

\end{document}